\documentstyle[12pt]{article}

\font\twlgot =eufm10 scaled \magstep1
\font\egtgot =eufm8
\font\sevgot =eufm7

\font\twlmsb =msbm10 scaled \magstep1
\font\egtmsb =msbm8
\font\sevmsb =msbm7

\newfam\gotfam
\def\pgot{\fam\gotfam\twlgot}
\textfont\gotfam\twlgot
\scriptfont\gotfam\egtgot
\scriptscriptfont\gotfam\sevgot
\def\got{\protect\pgot}

\newfam\msbfam
\textfont\msbfam\twlmsb
\scriptfont\msbfam\egtmsb
\scriptscriptfont\msbfam\sevmsb
\def\Bbb{\protect\pBbb}
\def\pBbb{\relax\ifmmode\expandafter\Bb\else\typeout{You cann't use
Bbb in text mode}\fi}
\def\Bb #1{{\fam\msbfam\relax#1}}

\def\thebibliography#1{\bigskip\section*{\large
\bf References\\}\list
  {[\arabic{enumi}]}{\settowidth\labelwidth{#1}\leftmargin\labelwidth
    \advance\leftmargin\labelsep
    \usecounter{enumi}}
    \def\newblock{\hskip .11em plus .33em minus .07em}
    \sloppy\clubpenalty4000\widowpenalty4000
    \sfcode`\.=1000\relax}

\def\op#1{\mathop{{\it\fam0} #1}\limits}

\newcommand{\id}{{\rm Id\,}}

\newcommand{\Ker}{{\rm Ker\,}}

\newcommand{\nm}[1]{\mid {#1}\mid}

\newcommand{\beq}{\begin{equation}}
\newcommand{\eeq}{\end{equation}}
\newcommand{\ben}{\begin{eqnarray}}
\newcommand{\een}{\end{eqnarray}}
\newcommand{\be}{\begin{eqnarray*}}
\newcommand{\ee}{\end{eqnarray*}}
\newcommand{\bea}{\begin{eqalph}}
\newcommand{\eea}{\end{eqalph}}

\newcommand{\gO}{{\got G}}
\newcommand{\cO}{{\cal O}}

\newcommand{\gj}{{\got J}}
\newcommand{\gE}{{\got E}}

\newcommand{\gQ}{{\got Q}}

\newcommand{\cL}{{\cal L}}

\newcommand{\cQ}{{\cal Q}}
\newcommand{\cE}{{\cal E}}

\newcommand{\bL}{{\bf L}}

\newcommand{\dl}{\delta}
\newcommand{\la}{\lambda}
\newcommand{\La}{\Lambda}
\newcommand{\f}{\phi}
\newcommand{\vf}{\varphi}

\newcommand{\om}{\omega}

\newcommand{\m}{\mu}

\newcommand{\G}{\Gamma}

\newcommand{\th}{\theta}

\newcommand{\si}{\sigma}
\newcommand{\Si}{\Sigma}

\newcommand{\w}{\wedge}

\newcommand{\ol}{\overline}

\newcommand{\dr}{\partial}

\newcommand{\ar}{\op\longrightarrow}

\newcommand{\ap}{\approx}

\newenvironment{eqalph}{\stepcounter{equation}
\setcounter{equationa}{\value{equation}}
\setcounter{equation}{0}

\begin{eqnarray}}{\end{eqnarray}\setcounter{equation}{\value{equationa}}}

\newcounter{example}
\newcounter{remark}
\newcounter{theorem}
\newcounter{proposition}
\newcounter{lemma}
\newcounter{corollary}
\newcounter{definition}

\def\thedefinition{\arabic{definition}}

\newenvironment{theo}{\refstepcounter{definition} \medskip\noindent{\bf
Theorem \thedefinition.}\sl}{\medskip }
\newenvironment{prop}{\refstepcounter{definition} \medskip\noindent{\bf
Proposition \thedefinition.}\sl}{\medskip }

\newenvironment{cor}{\refstepcounter{definition} \medskip\noindent{\bf 
Corollary \thedefinition.}\sl }{\medskip }

\newcommand{\mar}[1]{}

\begin{document}
\hbox{}

{\parindent=0pt

{\large \bf Noether conservation laws in infinite order Lagrangian
formalism} 
\bigskip 

{\bf G. Sardanashvily}

\medskip

\begin{small}

Department of Theoretical Physics, Moscow State University, 117234
Moscow, Russia

E-mail: sard@grav.phys.msu.su

URL: http://webcenter.ru/$\sim$sardan/
\bigskip

{\bf Abstract.}
Conservation laws related to the gauge invariance of Lagrangians and
Euler--Lagrange operators in finite and infinite order Lagrangian formalisms
are analyzed.
\end{small}
}

\section{Introduction}

Let us start from familiar finite order Lagrangian formalism.
Let $Y\to X$ be a smooth fibre bundle over an $n$-dimensional base $X$.
An $r$-order Lagrangian is defined as a density
\be
L:J^rY\op\ar_X \op\w^n T^*X 
\ee
on the $r$-order jet manifold $J^1Y$ of sections of $Y\to X$. Let $u$
be a vertical vector field on $Y\to X$ and $J^ru$ its prolongation
onto $J^rY\to X$. Let $\bL_{J^ru}L$ denote the Lie derivative of $L$
along $J^ru$. The first variational formula provides its
canonical decomposition
\mar{r1}\beq
\bL_{J^ru}L=u\rfloor\dl L + d_H(J^ru\rfloor H_L), \label{r1}
\eeq
where $\dl L$ is the Euler--Lagrange operator of
$L$, $d_H$ is the horizontal (total) differential (see (\ref{r11}) below)
and $H_L$ is a
Poincar\'e--Cartan form of $L$ (see \cite{got,krup3} for its explicit
expressions). Let $\bL_{J^ru}L$ vanishes everywhere on  
$J^rY$, i.e.,
a Lagrangian $L$ is invariant under a one-parameter group of vertical bundle
automorphisms (gauge transformations) of $Y\to X$ whose infinitesimal 
generator is $u$. Then, on Ker$\,\dl L$, one has the Noether conservation law
\mar{r2}\beq
0\ap d_H(J^ru\rfloor H_L) \label{r2}
\eeq
of the Noether current 
\mar{r3}\beq
\gj_u=J^ru\rfloor H_L. \label{r3}
\eeq
Note that, unless $r\leq 2$, a Poincar\'e--Cartan form $H_L$ is not unique.
Moreover, one can put $\gj_u=h_0(J^{2r-1}u\rfloor\rho_L)$ where $h_0$ is the
horizontal projection (see (\ref{r10}) below)
and $\rho_L$ is an arbitrary Lepagean equivalent
of a Lagrangian $L$.

However, it may happen that, though the Lie derivative $\bL_{J^ru}L$ does not
vanish, a conservation law takes place. Indeed, let this Lie derivative be
a horizontal differential 
\mar{r4}\beq
\bL_{J^ru}L=d_H\si. \label{r4}
\eeq
Then, the first variational formula (\ref{r1}) on Ker$\,\dl L$ leads to
the equality
\mar{r5}\beq
0\ap d_H(\gj_u-\si), \label{r5}
\eeq
regarded as a conservation law of the modified Noether current $\gj_u-\si$.

In order to understand the condition (\ref{r4}), let us refer to the 
master identity
\mar{r6}\beq
\dl(\bL_{J^ru}L)=\bL_{J^{2r}u}(\dl L) \label{r6}
\eeq
(see Appendix). It follows that the Euler--Lagrange
operator is invariant under a one-parameter gauge group generated by
$u$ iff the Lie derivative $\bL_{J^ru}L$ is a variationally trivial
Lagrangian. The Lie derivative (\ref{r4}) is such a Lagrangian as follows. 

\begin{theo} \label{tt1} \mar{tt1}
An $r$-order Lagrangian $L$ (\ref{r1}) is variationally trivial iff it
takes the form
\mar{r7}\beq
L=d_H\xi +h_0\vf \label{r7}
\eeq
where $\xi$ is an $n-1$-form of jet order $r-1$ and $\vf$ is a closed
$n$-form on $Y$. 
\end{theo}

This assertion has been proved by a computation of cohomology of finite
order variational sequences \cite{and,bor,krup1,krup2,vit}. It is also
reproduced by a computation of cohomology of the infinite order
variational complex, but without minimizing the jet order of the form
$\xi$ \cite{lmp,jmp,ijmms,epr,vin}.

\begin{cor} \label{tt2} \mar{tt2}
It follows from the master identity (\ref{r6}) and Theorem \ref{tt1} that
the Euler--Lagrange operator $\dl L$ of a Lagrangian $L$ is 
invariant under a one-parameter group of gauge transformations
generated by a vector field $u$ iff the Lie derivative $\bL_{J^ru}L$
of this Lagrangian takes the form 
\mar{r8}\beq
\bL_{J^ru}L=d_H\si +h_0\f \label{r8}
\eeq
where $\f$ is a closed $n$-form on $Y$. 
\end{cor}

The equality (\ref{r8}) locally reduces to the equality (\ref{r4})
known as the Noether--Bessel--Hagen equation \cite{krup3}. If the
equality (\ref{r8}) globally takes the form (\ref{r4}), the
conservation law (\ref{r5}) holds.

The Lie derivative of a global Chern--Simons Lagrangian 
illustrates the formula (\ref{r8}) \cite{bor2}.

A differential operator $\cE$ on $Y\to X$ is said to be locally variational if 
each point of $Y$ admits an open neighbourhood such that, on this
neighbourhood, $\cE$ is the Euler--Lagrange operator of some local Lagrangian.

\begin{theo} \label{tt3} \mar{tt3}
A $2k$-order differential operator $\cE$ is locally variational iff
\mar{r9}\beq
\cE=\dl L + \tau(\vf), \label{r9}
\eeq
where $L$ is a $k$-order Lagrangian, $\tau$ is a certain differential operator 
such that $\dl=\tau\circ d$
(see (\ref{r12}) below) and $\vf$ is a closed
$(n+1)$-form on $Y$. 
\end{theo}

For instance, if $Y\to X$ is an affine bundle, its de Rham cohomology
equals that of 
$X$ and, consequently, any variationally trivial operator on $Y$ is the
Euler--Lagrange operator of some global Lagrangian. Then, Corollary
\ref{tt2} can be applied to this operator. The above mentioned global
Chern--Simons model illustrates this fact.

Theorem \ref{tt3} gives a solution of the global inverse problem in 
finite order Lagrangian formalism \cite{and} (see also
\cite{bor,krup1,krup3,vit}). This Theorem as like as Theorem \ref{tt1}
issues from a computation of cohomology of the infinite variational
complex, but without minimizing the order of a Lagrangian $L$
\cite{lmp,jmp,ijmms,epr}.
Infinite order jet formalism and the infinite variational complex is a
convenient tool of studying Lagrangian systems both of infinite and 
finite order (see, e.g.,
\cite{book,epr}). Note that infinite order jets 
are also utilized in some
quantum field models \cite{barn,catt,fulp,lmp,mpla}.
Our goal here is the extension of the first variational formula (\ref{r1}), 
the Noether conservation law (\ref{r2}) and the master identity
(\ref{r6}) to infinite
order Lagrangians.

\section{The differential calculus in infinite order jets}

Smooth manifolds throughout are assumed to be
real, finite-dimensional, Hausdorff, second-countable (i.e., paracompact), 
and connected.  
We follow the terminology of
\cite{bred,hir}, where a sheaf $S$ is a particular topological bundle,
$\ol S$ denotes the canonical presheaf of
sections of the sheaf $S$, and 
$\G(S)$ is the group of global sections of $S$. 

Recall that the infinite order jet space  of a smooth fibre bundle
$Y\to X$ is defined as a projective limit $(J^\infty Y,\{\pi^\infty_r\})$ 
of the inverse system
\mar{t1}\beq
X\op\longleftarrow^\pi Y\op\longleftarrow^{\pi^1_0}\cdots \longleftarrow
J^{r-1}Y \op\longleftarrow^{\pi^r_{r-1}} J^rY\longleftarrow\cdots \label{t1}
\eeq
of finite order jet manifolds $J^rY$ of $Y\to X$. 
Endowed with the projective limit topology,
$J^\infty Y$ is a paracompact Fr\'echet manifold \cite{tak2}.
A bundle coordinate atlas
$\{U,(x^\la,y^i)\}$ of $Y\to X$ yields the manifold
coordinate atlas
\be
\{(\pi^\infty_0)^{-1}(U), (x^\la, y^i_\La)\}, \qquad 0\leq|\La|,
\ee
 of $J^\infty
Y$, together with the transition functions  
\be
{y'}^i_{\la+\La}=\frac{\dr x^\m}{\dr x'^\la}d_\m y'^i_\La, 
\ee
where $\La=(\la_k\ldots\la_1)$, $\la+\La=(\la\la_k\ldots\la_1)$ are
multi-indices and
\be
d_\la = \dr_\la + \op\sum_{|\La|\geq 0} y^i_{\la+\La}\dr_i^\La
\ee
is the total derivative.
We will also use the notation $d_\La=d_{\la_k}\cdots d_{\la_1}$, $\La=
(\la_k\ldots\la_1)$.

With the inverse system (\ref{t1}), one has
the direct system 
\be
\cO^*(X)\op\longrightarrow^{\pi^*} \cO^*_0 
\op\longrightarrow^{\pi^1_0{}^*} \cO_1^*
\op\longrightarrow^{\pi^2_1{}^*} \cdots \op\longrightarrow^{\pi^r_{r-1}{}^*}
 \cO_r^* \longrightarrow\cdots 
\ee
of graded differential $\Bbb R$-algebras $\cO^*_r$ of exterior forms on finite
order jet manifolds $J^rY$, where $\pi^r_{r-1}{}^*$ are the pull-back
monomorphisms. The direct limit of this direct system is the graded 
differential algebra $\cO^*_\infty$ 
of exterior forms on  finite order jet manifolds modulo the pull-back
identification. However, $\cO^*_\infty$ does not exhaust all exterior forms
on $J^\infty Y$.

Let $\gO^*_r$ be a sheaf
of germs of exterior forms on the $r$-order jet manifold $J^rY$ and 
$\ol\gO^*_r$ its canonical presheaf.  There is the direct system of canonical
presheaves
\be
\ol\gO^*_X\op\longrightarrow^{\pi^*} \ol\gO^*_0 
\op\longrightarrow^{\pi^1_0{}^*} \ol\gO_1^*
\op\longrightarrow^{\pi^2_1{}^*} \cdots \op\longrightarrow^{\pi^r_{r-1}{}^*}
 \ol\gO_r^* \longrightarrow\cdots, 
\ee
where $\pi^r_{r-1}{}^*$ are the pull-back monomorphisms. Its direct
limit $\ol\gO^*_\infty$ 
is a presheaf of graded differential
$\Bbb R$-algebras on
$J^\infty Y$. Let $\gQ^*_\infty$ be a sheaf constructed from 
$\ol\gO^*_\infty$, $\ol\gQ^*_\infty$ its canonical presheaf, 
and $\cQ^*_\infty=\G(\gQ^*_\infty)$ the structure algebra of 
sections of
the sheaf $\gQ^*_\infty$. 
There are 
$\Bbb R$-algebra monomorphisms 
 $\ol\gO^*_\infty
\to\ol\gQ^*_\infty$ and  $\cO^*_\infty
\to\cQ^*_\infty$. 

The key point is that, since the paracompact space
$J^\infty Y$ admits a partition of unity by elements of the ring
$\cQ^0_\infty$ \cite{tak2}, the sheaves of
$\cQ^0_\infty$-modules on
$J^\infty Y$ are fine and, consequently, acyclic. Therefore, the 
abstract de Rham theorem on cohomology of a sheaf resolution \cite{hir}
can be called into play in order to obtain cohomology of the graded
differential algebra $\cQ^*_\infty$. In turn, $\cO^*_\infty$
is proved to possess the same cohomology as $\cQ^*_\infty$
(see Theorem \ref{am11} below) \cite{lmp,jmp,ijmms,epr}. 

For short, we
agree to call elements of $\cQ^*_\infty$ the
exterior forms on
$J^\infty Y$.  Restricted to a
coordinate chart
$(\pi^\infty_0)^{-1}(U_Y)$ of $J^\infty Y$, they
can be written in a coordinate form, where horizontal forms
$\{dx^\la\}$ and contact 1-forms
$\{\th^i_\La=dy^i_\La -y^i_{\la+\La}dx^\la\}$ provide local
generators of the algebra
$\cQ^*_\infty$. 
There is the canonical decomposition
\be
\cQ^*_\infty =\op\oplus_{k,s}\cQ^{k,s}_\infty, \qquad 0\leq k, \qquad
0\leq s\leq n,
\ee
of $\cQ^*_\infty$ into $\cQ^0_\infty$-modules $\cQ^{k,s}_\infty$
of $k$-contact and $s$-horizontal forms, together with the corresponding
projections
\mar{r10}\beq
h_k:\cQ^*_\infty\to \cQ^{k,*}_\infty, \quad 0\leq k, \qquad
h^s:\cQ^*_\infty\to \cQ^{*,s}_\infty, \quad 0\leq s
\leq n. \label{r10}
\eeq
Accordingly, the
exterior differential on $\cQ_\infty^*$ is split
into the sum $d=d_H+d_V$ of horizontal and vertical
differentials such that
\mar{r11}\ben
&& d_H\circ h_k=h_k\circ d\circ h_k, \qquad d_H(\f)=
dx^\la\w d_\la(\f), \label{r11}\\ 
&& d_V \circ h^s=h^s\circ d\circ h^s, \qquad
d_V(\f)=\th^i_\La \w \dr^\La_i\f, \qquad \f\in\cQ^*_\infty. \nonumber
\een

\section{The infinite variational complex}

Being nilpotent, the
differentials $d_V$ and $d_H$ provide the natural bicomplex
$\{\gQ^{k,m}_\infty\}$ of  the sheaf
$\gQ^*_\infty$ on $J^\infty Y$. To complete it to the
variational bicomplex, one defines the projection $\Bbb R$-module
endomorphism 
\mar{r12}\ben
&& \tau=\op\sum_{k>0} \frac1k\ol\tau\circ h_k\circ h^n, \label{r12}\\ 
&&\ol\tau(\f)= \op\sum_{|\La|\geq 0}
(-1)^{\nm\La}\th^i\w [d_\La(\dr^\La_i\rfloor\f)], 
\qquad \f\in \ol\gO^{>0,n}_\infty, \nonumber
\een
of $\ol\gO^*_\infty$ such that
\be
\tau\circ d_H=0, \qquad  \tau\circ d\circ \tau -\tau\circ d=0.
\ee
Introduced on elements of the presheaf $\ol\gO^*_\infty$ 
(see, e.g., \cite{bau,book,tul}), this endomorphism is induced on the
sheaf $\gQ^*_\infty$ and its structure algebra
$\cQ^*_\infty$. Put
\be
\gE_k=\tau(\gQ^{k,n}_\infty), \qquad E_k=\tau(\cQ^{k,n}_\infty), \qquad k>0.
\ee
Since
$\tau$ is a projection operator, we have isomorphisms 
\be
\ol\gE_k=\tau(\ol\gQ^{k,n}_\infty), \qquad E_k=\G(\gE_k).
\ee
The variational operator on $\gQ^{*,n}_\infty$ is defined as the
morphism $\dl=\tau\circ d$. 
It is nilpotent, and obeys the relation 
\mar{am13}\beq
\dl\circ\tau-\tau\circ d=0. \label{am13}
\eeq

Let $\Bbb R$ and  $\gO^*_X$ denote the constant sheaf
on
$J^\infty Y$ and the sheaf of exterior forms on $X$, respectively. The
operators $d_V$,
$d_H$,
$\tau$ and $\dl$ give the following variational bicomplex of
sheaves of differential forms on $J^\infty Y$:
\mar{7}\beq
\begin{array}{ccccrlcrlccrlccrlcrl}
& & & & _{d_V} & \put(0,-7){\vector(0,1){14}} & & _{d_V} &
\put(0,-7){\vector(0,1){14}} & &  & _{d_V} &
\put(0,-7){\vector(0,1){14}} & & &  _{d_V} &
\put(0,-7){\vector(0,1){14}}& & _{-\dl} & \put(0,-7){\vector(0,1){14}} \\ 
 &  & 0 & \to & &\gQ^{k,0}_\infty &\ar^{d_H} & & \gQ^{k,1}_\infty &
\ar^{d_H} &\cdots  & & \gQ^{k,m}_\infty &\ar^{d_H} &\cdots & &
\gQ^{k,n}_\infty &\ar^\tau &  & \gE_k\to  0\\  
 & &  &  & & \vdots & & & \vdots  & & & 
&\vdots  & & & &
\vdots & &   & \vdots \\ 
& & & & _{d_V} & \put(0,-7){\vector(0,1){14}} & & _{d_V} &
\put(0,-7){\vector(0,1){14}} & &  & _{d_V} &
\put(0,-7){\vector(0,1){14}} & & &  _{d_V} &
\put(0,-7){\vector(0,1){14}}& & _{-\dl} & \put(0,-7){\vector(0,1){14}} \\ 
 &  & 0 & \to & &\gQ^{1,0}_\infty &\ar^{d_H} & & \gQ^{1,1}_\infty &
\ar^{d_H} &\cdots  & & \gQ^{1,m}_\infty &\ar^{d_H} &\cdots & &
\gQ^{1,n}_\infty &\ar^\tau &  & \gE_1\to  0\\  
& & & & _{d_V} &\put(0,-7){\vector(0,1){14}} & & _{d_V} &
\put(0,-7){\vector(0,1){14}} & & &  _{d_V}
 & \put(0,-7){\vector(0,1){14}} & &  & _{d_V} & \put(0,-7){\vector(0,1){14}}
 & & _{-\dl} & \put(0,-7){\vector(0,1){14}} \\
0 & \to & \Bbb R & \to & & \gQ^0_\infty &\ar^{d_H} & & \gQ^{0,1}_\infty &
\ar^{d_H} &\cdots  & &
\gQ^{0,m}_\infty & \ar^{d_H} & \cdots & &
\gQ^{0,n}_\infty & \equiv &  & \gQ^{0,n}_\infty \\
& & & & _{\pi^{\infty*}}& \put(0,-7){\vector(0,1){14}} & & _{\pi^{\infty*}} &
\put(0,-7){\vector(0,1){14}} & & &  _{\pi^{\infty*}}
 & \put(0,-7){\vector(0,1){14}} & &  & _{\pi^{\infty*}} &
\put(0,-7){\vector(0,1){14}} & &  & \\
0 & \to & \Bbb R & \to & & \gO^0_X &\ar^d & & \gO^1_X &
\ar^d &\cdots  & &
\gO^m_X & \ar^d & \cdots & &
\gO^n_X & \ar^d & 0 &  \\
& & & & &\put(0,-5){\vector(0,1){10}} & & &
\put(0,-5){\vector(0,1){10}} & & & 
 & \put(0,-5){\vector(0,1){10}} & & &   &
\put(0,-5){\vector(0,1){10}} & &  & \\
& & & & &0 & &  & 0 & & & & 0 & & & & 0 & &  & 
\end{array}
\label{7}
\eeq
The second row and the last column of this bicomplex assemble into the 
infinite variational complex
\mar{tams1}\beq
0\to\Bbb R\to \gQ^0_\infty \ar^{d_H}\gQ^{0,1}_\infty\ar^{d_H}\cdots  
\op\longrightarrow^{d_H} 
\gQ^{0,n}_\infty  \op\longrightarrow^\dl \gE_1 
\op\longrightarrow^\dl 
\gE_2 \longrightarrow \cdots\, . \label{tams1}
\eeq
The corresponding variational bicomplex and variational complex of
the graded differential algebra 
$\cQ^*_\infty$ (see (\ref{b317}) below) take place.

There are the well-known statements summarized usually as
the algebraic Poincar\'e lemma (see, e.g., \cite{olver,tul}). 

\begin{theo} \label{am12} \mar{am12}
If $Y$ is a contractible bundle $\Bbb R^{n+p}\to \Bbb R^n$, the
variational bicomplex of the graded differential algebra $\cO^*_\infty$
is exact.
\end{theo}

It follows that the variational bicomplex (\ref{7}) and, consequently,
the variational complex (\ref{tams1}) are exact for any smooth bundle
$Y\to X$.
Moreover, the sheaves
$\gQ^{k,m}_\infty$ and $\gE_k$ are fine.
Thus, the columns and rows of the bicomplex (\ref{7}) as like as the
variational complex (\ref{tams1}) are sheaf resolutions, and the
abstract de Rham theorem can be applied to them. 
The results are the following \cite{and,ander,jmp,ijmms,epr,tak2}.

 Let us start from the following assertion.

\begin{prop} \label{20jpa} \mar{20jpa}
Since $Y$ is a strong deformation retract of $J^\infty Y$, there is an
isomorphism 
\mar{lmp80}\beq
H^*(J^\infty Y,\Bbb R)= H^*(Y,\Bbb R)=H^*(Y) \label{lmp80}
\eeq
between cohomology $H^*(J^\infty Y,\Bbb R)$ of $J^\infty Y$ with
coefficients in the constant sheaf $\Bbb R$, that $H^*(Y,\Bbb R)$ of $Y$, and
the de Rham cohomology $H^*(Y)$ of $Y$. 
\end{prop}

Let us consider the de Rham complex of sheaves 
\mar{lmp71}\beq
0\to \Bbb R\to
\gQ^0_\infty\op\longrightarrow^d\gQ^1_\infty\op\longrightarrow^d
\cdots
\label{lmp71}
 \eeq
on $J^\infty Y$ and the corresponding de Rham complex of their structure
algebras
\mar{5.13'}\beq
0\to \Bbb R\to
\cQ^0_\infty\op\longrightarrow^d\cQ^1_\infty\op\longrightarrow^d
\cdots\, .
\label{5.13'}
\eeq
The complex (\ref{lmp71}) is exact due to
the Poincar\'e lemma, and is a resolution of the constant sheaf $\Bbb R$ on
$J^\infty Y$ since sheaves $\gQ^r_\infty$ are fine. Then, the abstract de
Rham theorem and Lemma
\ref{20jpa} lead to the following.

\begin{prop} \label{38jp} \mar{38jp}
The de Rham cohomology $H^*(\cQ^*_\infty)$ 
of the graded differential algebra
$\cQ^*_\infty$  is isomorphic to that $H^*(Y)$ of the bundle $Y$.
\end{prop}

It follows that every closed form $\f\in \cQ^*_\infty$
is split into the sum
\mar{tams2}\beq
\f=\varphi +d\xi, \qquad \xi\in \cQ^*_\infty, \label{tams2} 
\eeq
where $\varphi$ is a closed form on the fibre bundle $Y$. 

Turn now to the rows of the variational bicomplex (\ref{7}). We have
the exact sequence of sheaves
\be
0\to \gQ^{k,0}_\infty\ar^{d_H} \gQ^{k,1}_\infty\ar^{d_H}\cdots
\ar^{d_H}\gQ^{k,n}_\infty\ar^\tau\gE_k\to 0, \qquad k>0.
\ee
Since the sheaves $\gQ^{k,m}_\infty$ and $\gE_k$ are fine, this is a
resolution of the fine sheaf $\gQ^{k,0}_\infty$. Then, the abstract de
Rham theorem results in the following.

\begin{prop} \label{tt5} \mar{tt5}
The cohomology groups of the complex
\mar{r20}\beq
0\to \cQ^{k,0}_\infty\ar^{d_H} \cQ^{k,1}_\infty\ar^{d_H}\cdots
\ar^{d_H}\cQ^{k,n}_\infty\ar^\tau E_k\to 0, \qquad k>0, \label{r20}
\eeq
are trivial.
\end{prop}

The variational complex (\ref{tams1}) is a resolution of the constant
sheaf $\Bbb R$ on $J^\infty Y$.
Then, from the abstract de Rham theorem and Proposition 
\ref{20jpa}, we obtain the following. 

\begin{prop} \label{lmp05} \mar{lmp05}
There is an isomorphism
between $d_H$- and $\dl$-cohomology of the
variational complex 
\mar{b317}\beq
0\to\Bbb R\to \cQ^0_\infty \ar^{d_H}\cQ^{0,1}_\infty\ar^{d_H}\cdots  
\op\longrightarrow^{d_H} 
\cQ^{0,n}_\infty  \op\longrightarrow^\dl E_1 
\op\longrightarrow^\dl 
E_2 \longrightarrow \cdots  \label{b317}
\eeq
and the de Rham cohomology of the fibre bundle
$Y$, namely,
\be
H^{k<n}(d_H;\cQ^*_\infty)=H^{k<n}(Y), \qquad H^{k-n}(\dl;
\cQ^*_\infty)=H^{k\geq n}(Y).
\ee
\end{prop}

Moreover, the relation (\ref{am13}) for $\tau$ and
the relation $h_0d=d_Hh_0$ for $h_0$ define  a homomorphisms of the
de Rham complex (\ref{5.13'}) of the algebra $\cQ^*_\infty$ to its variational
complex (\ref{b317}). The corresponding homomorphism of their cohomology
groups is an isomorphism by virtue of Proposition \ref{38jp} and Proposition
\ref{lmp05}. Then, the splitting (\ref{tams2}) leads to the following
decompositions.

\begin{theo} \label{t41} \mar{t41}
Any $d_H$-closed form $\si\in\cQ^{0,m}$, $m< n$, is represented by a sum
\beq
\si=h_0\varphi+ d_H \xi, \qquad \xi\in \cQ^{m-1}_\infty, \label{t60}
\eeq
where $\varphi$ is a closed $m$-form on $Y$.
Any $\dl$-closed form $\psi\in\cQ^{k,n}$, $k\geq 0$, is split into
\mar{t42a-c}\ben
&& \psi=h_0\varphi + d_H\xi, \qquad k=0, \qquad \xi\in \cQ^{0,n-1}_\infty,
\label{t42a}\\ 
&& \psi=\tau(\varphi) +\dl(\xi), \qquad k=1, \qquad \xi\in \cQ^{0,n}_\infty,
\label{t42b}\\
&& \psi=\tau(\varphi) +\dl(\xi), \qquad k>1, \qquad \xi\in E_{k-1},
\label{t42c}
\een
where $\varphi$ is a closed $(n+k)$-form on $Y$. 
\end{theo}
 
The variational complex (\ref{b317}) provides the algebraic approach to
the calculus of variations in the class of exterior forms of locally
finite jet order \cite{bau,book,tul}. For instance, the variational
operator $\dl$ acting on $\cQ^{0,n}_\infty$ is the Euler--Lagrange map,
while $\dl$ acting on $E_1$ is the Helmholtz--Sonin map. Accordingly,
one can think of a horizontal density
\be
L=\cL\om, \qquad \om=dx^1\w\cdots\w dx^n,
\ee
on $J^\infty Y$ as being a Lagrangian of locally finite order. 
Then, the expressions (\ref{t42a}) -- (\ref{t42b}) in Theorem \ref{t41} 
give a
solution of the global inverse problem of the calculus of variations on fibre
bundles in the class of Lagrangians $L\in\cQ^{0,n}_\infty$ of locally finite
order. Namely, a Lagrangian $L\in \cQ^{0,n}_\infty$ is variationally trivial
iff it takes the form (\ref{t42a}), while an
Euler--Lagrange-type operator $\cE\in E_1$ satisfies the Helmholtz
condition $\dl(\cE)=0$ iff it takes the form (\ref{t42b}).

In order to return to Theorems \ref{tt1} and \ref{tt3}, let us consider
the subalgebra $\cO^*_\infty\subset \cQ^*_\infty$ of exterior forms of
bounded jet order. It makes up a subcomplex of the variational complex
(\ref{b317}). The key point is the following \cite{lmp,jmp,ijmms,epr}.

\begin{theo} \label{am11} \mar{am11}
Graded differential algebra $\cO^*_\infty$ has the same $d-$, $d_H$- and
$\dl$-cohomology as $\cQ^*_\infty$.
\end{theo}

It follows that, if an exterior forms $\psi$ in the formulas (\ref{t60}) --
(\ref{t42c}) are of finite jet order, then the exterior form $\xi$ are so.
In particular, we come to Theorems \ref{tt1} and \ref{tt3}, but without
minimizing the jet order of the exterior forms $\xi$ and $L$, respectively.

\section{Conservation laws}

The exactness of the complex (\ref{r20}) at the term $\cQ^{k,n}_\infty$
implies that, if $\tau(\f)=0$, $\f\in \cQ^{k,n}_\infty$, then
$\f=d_H\xi$, $\xi\in \cQ^{k,n-1}_\infty$. Since $\tau$ is a projection
operator, there is the $\Bbb R$-module decomposition
\mar{30jpa}\beq
\cQ^{k,n}_\infty=E_k\oplus d_H(\cQ^{k,n-1}_\infty). \label{30jpa}
\eeq
Given a Lagrangian $L\in \cQ^{0,n}_\infty$, the decomposition (\ref{30jpa})
in the case of $k=1$ reads
\mar{+421}\beq
dL=\tau(dL) + (\id -\tau)(dL)= \dl L + d_H(\phi), \label{+421}
\eeq
where $\phi\in \cQ^{1,n-1}_\infty$ and
\mar{r26}\beq
\dl L= \cE_i\th^i\w\om= \op\sum_{|\La|\geq 0}
(-1)^{\mid \La\mid} d_\La (\dr^\La_i\cL) \th^i\w\om \label{r26}
\eeq
is the Euler--Lagrange operator of an infinite order Lagrangian $L$. 

Let $u=u^i\dr_i$ be a vertical vector  field on a fibre bundle $Y\to
X$ seen as a 
generator of one-parameter gauge group. It
defines the derivation
\mar{r27}\beq
J^\infty u=\op\sum_{|\La|\geq 0}d_\La u^i\dr_i^\La  \label{r27}
\eeq
of the ring $\cQ^0_\infty$ regarded as an
infinite order jet prolongation of $u$ onto $J^\infty Y$.
We also have the
contraction $u\rfloor\f$ and the  Lie derivative 
\be
\bL_{J^\infty u}\f= J^\infty u\rfloor d\f + d(J^\infty u\rfloor
\f)
\ee 
of elements of the differential algebra $\cQ^*_\infty$. It is easily
justified that
\be
J^\infty u\rfloor d_H\f=- d_H(J^\infty u\rfloor\f), \qquad \f\in
\cQ^*_\infty. 
\ee

Let $L\in \cQ^{0,n}_\infty$ be an infinite order Lagrangian. 
By virtue of the decomposition (\ref{+421}), we come to the first
variational formula 
\mar{hn7}\beq
\bL_{J^\infty u}L= J^\infty u\rfloor dL= u\rfloor \dl L - d_H (J^\infty
u\rfloor\phi),
\label{hn7}
\eeq
where 
\be
\gj_u=- J^\infty u\rfloor\phi \in \cQ^{0,n-1}_\infty 
\ee
is the symmetry current along the vector field $u$. 
If $L$ is a finite order Lagrangian, this current is given by the expression  
(\ref{r3}) modulo a $d_H$-closed form. However, a glance at
the explicit formulas for Lepagean equivalents \cite{got,krup3} shows
that this expression can not be generalized to the case of infinite order
Lagrangians. 
If the Lie derivative $\bL_{J^\infty u}L$ 
vanishes, the first variational formula (\ref{hn7}) leads to the Noether
conservation law
\mar{r23}\beq
d_HJ_u\ap 0 \label{r23}
\eeq
on Ker$\,\dl L$, i.e., the global section $d_HJ_u$ of the sheaf
$\gQ^{0,n}_\infty$ on $J^\infty Y$ takes zero values at points of the
subspace Ker$\,\dl L\subset J^\infty Y$ given by the condition $\dl L=0$.

There is the master identity 
\mar{r22}\beq
\dl(\bL_{J^\infty u}L)=\bL_{J^\infty u}(\dl L) \label{r22}
\eeq
(see Appendix for its proof). It follows from this identity and Theorem
\ref{t41} that 
the Euler--Lagrange operator $\dl L$ (\ref{r26}) of an infinite order
Lagrangian $L$ is  
invariant under a one-parameter group of gauge transformations
generated by a vector field $u$ iff the Lie derivative $\bL_{J^\infty u}L$
of this Lagrangian takes the form 
\be
\bL_{J^\infty u}L=d_H\si +h_0\f,
\ee
where $\f$ is a closed $n$-form on $Y$. 

In conclusion, let us say a few words on the cohomology of conservation
laws in infinite (and finite) order jet formalism.
If the conservation law (\ref{r23}) takes place, 
one can say that the horizontal differential $d_HJ_u$ is a relative
$d_H$-cocycle on the pair of topological spaces $(J^\infty Y,\Ker \dl L)$.
Of course, it is a
$d_H$-coboundary, but need not be a relative
$d_H$-coboundary since $J_u\not\ap 0$. Therefore,
the horizontal differential $d_HJ_u$ of a conserved current $J_u$ can
be characterized by 
elements of the relative $d_H$-cohomology group
$H^n_{\rm rel}(J^\infty Y,\Ker
\dl L)$ of the pair $(J^\infty Y,\Ker
\dl L)$.

For
instance, any conserved Noether current in the Yang--Mills gauge theory on a
principal bundle $P$ with a structure group $G$ is well known to reduce to a
superpotential, i.e.,
$J_u=W+d_H U$ where
$W\ap 0$. Its horizontal differential $d_HJ_u$ belongs to the trivial
element of 
the relative cohomology group $H^n_{\rm rel}(J^2 Y,\Ker\dl L_{\rm YM})$,
where $Y=J^1P/G$.

Let now $N^n\subset X$ be an $n$-dimensional submanifold of $X$ with a
compact boundary $\dr N^n$. Let $s$ be a section of the fibre bundle 
$Y\to X$ and $\ol s=J^\infty s$ its infinite order jet prolongation, i.e.,
$y^i_\La\circ \ol s=d_\La s^i$, $0<|\La|$. Let us assume that $\ol s(\dr N^n)
\subset \Ker\dl L$. Then, the quantity
\mar{35jp}\beq
\op\int_{N^n} \ol s^*d_HJ_u= \op\int_{\dr N^n} \ol s^*J_u \label{35jp}
\eeq
depends only on the relative cohomology class of the divergence $d_HJ_u$.
For instance, in the above mentioned case of gauge theory, the quantity
(\ref{35jp}) vanishes. 

Let $N^{n-1}$ be a compact $(n-1)$-dimensional submanifold of $X$ without
boundary, and $s$ a section of $Y\to X$ such that $\ol
s(N^{n-1})\subset\Ker\dl L$.
Let $J_u$ and $J'_u$ be two currents in the first variational formula
(\ref{hn7}). They differ from each
other in a $d_H$-closed form $\varphi$. Then, the difference
\mar{36jp}\beq
\op\int_{N^{n-1}}\ol s^*(J_u - J'_u) \label{36jp}
\eeq
depends only on the
homology class of $N^{n-1}$ and the de Rham cohomology class of $\ol
s^*\varphi$. The latter is an image of the $d_H$-cohomology class of $\varphi$
under the morphisms
\be
H^{n-1}(d_H)\ar^{h_0} H^{n-1}(Y) \ar^{s^*} H^{n-1}(X).
\ee
In particular, if $N^{n-1}=\dr N^n$ is a boundary, the quantity (\ref{36jp})
always vanishes.

\section{Appendix}

In order to prove the master identity (\ref{r22}), let us act on the 
first variational formula (\ref{hn7}) by the variational operator $\dl$.
Since $\dl\circ d_H=0$, we obtain the equality
\be
\dl(\bL_{J^\infty u}L)= \dl(u\rfloor \dl L).
\ee
Therefore, we aim to prove that
\mar{r25}\beq
\dl(u\rfloor \dl L)=\bL_{J^\infty u}\dl L. \label{r25}
\eeq

It suffices to show that, given an arbitrary point $y\in Y$, there
exists its open neighbourhood $U$ such that the equality (\ref{r25})
holds on $(\pi^\infty_0)^{-1}(U)$. Using the coordinate expressions
(\ref{r26}) -- (\ref{r27}), let us write
\be
&& \dl(u\rfloor \dl L)=\dl(u^i\cE_i\om)=[\dr_k(u^i\cE_i)+ \op\sum_{|\La|> 0}
(-1)^{|\La|}d_\La\dr^\La_k(u^i\cE_i)] dy^k\w\om, \\
&& \bL_{J^\infty u}\dl L=d(u^i\cE_i)\w\om +J^\infty u\rfloor d(\dl L)=\\
&& \qquad \dr_k(u^i\cE_i)dy^k\w\om+ u\rfloor(\dr_k\cE_i dy^k\w dy^i\w\om) +
\op\sum_{|\La|> 0} (d_\La u^i)\dr^\La_i\cE_k dy^k\w\om.
\ee
Then, the equality (\ref{r25}) takes the form
\mar{r30}\beq
\op\sum_{|\La|> 0}
(-1)^{|\La|}d_\La\dr^\La_k(u^i\cE_i)dy^k\w\om=
u\rfloor(\dr_k\cE_i dy^k\w dy^i\w\om) +
\op\sum_{|\La|> 0} (d_\La u^i)\dr^\La_i\cE_k dy^k\w\om.
 \label{r30}
\eeq

Let us further assume that $u(y)\neq 0$. In this case, there exists an
open neighbourhood $U$ of $y$ provided with bundle coordinates $(x^\la,y'^i)$
such that $u=\dr_1$. With respect to these coordinates, the equality
(\ref{r30}) reads
\be
\op\sum_{|\La|> 0}
(-1)^{|\La|}d_\La\dr^\La_k(\op\sum_{|\Si|\geq 0}
(-1)^{|\Si|}d_\Si\dr^\Si_1\cL)dy^k\w\om=(\dr_1\cE_k-\dr_k\cE_1)dy^k\w\om.
\ee
It is brought into the form
\be
\dl((\cE_1-\dr_1\cL)\om)=0.
\ee
This equality really holds since $(\cE_1-\dr_1\cL)\om$ is a
variationally trivial Lagrangian due to the first variational formula
(\ref{hn7}) where $u=\dr_1$, i.e.,
\be
\dr_1\cL\om=\cE_1\om - d_H (\phi_1).
\ee

If $u(y)=0$. There exists a vertical vector field $u'$ such that
$u'(y)\neq 0$. The equality (\ref{r25}) holds both for $u'$ and $u+u'$
and, consequently, does so for $u$.

If $L$ is a finite order Lagrangian, we obtain the master identity
(\ref{r6}).


\begin{thebibliography}{ederf}

\bibitem{and} I.Anderson and T.Duchamp, On the existence of global
variational principles, {\it Amer. J. Math.} {\bf 102} (1980) 781.

\bibitem{ander} I.Anderson, Introduction to the variational bicomplex,
{\it Contemp. Math.} {\bf 132} (1992) 51. 

\bibitem{barn} G.Barnish, F.Brandt and M.Henneaux, Local BRST
cohomology in gauge theories, {\it Phys. Rep.} {\bf 338} (2000) 439.

\bibitem{bau} M.Bauderon, Differential geometry and Lagrangian formalism
in the calculus of variations, In: {\it Differential Geometry, Calculus of
Variations, and their Applications}, Lecture Notes in Pure and Applied
Mathematics {\bf 100} (Marcel Dekker Inc., 
New York, 1985) 67.

\bibitem{bor} A.Borowiec, M.Ferraris, M.Francaviglia and M.Palese,
Conservation laws for non-global Lagrangians, {\it E-print arXiv}:
math-ph/0301043. 

\bibitem{bor2} A.Borowiec, M.Ferraris and M.Francaviglia, A covariant
formalism for Chern--Simons gravity, {\it E-print arXiv}:
math-ph/0301146.

\bibitem{bred} G. Bredon, {\it Sheaf Theory} (McGraw-Hill Book Company,
New York, 1967).  

\bibitem{catt} A.Cattaneo, G.Felder and L.Tomassini, Fedosov
connections on jet bundles and deformation quantization, 
{\it E-print arXiv}: math.QA/0111290.

\bibitem{fulp} R.Fulp, T.Lada and J.Stasheff, Noether's variational
theorem II and the BV formalism, {\it E-print arXiv}: math.QA/0204079.

\bibitem{book} G.Giachetta, L. Mangiarotti and G. Sardanashvily, {\it
Lagrangian and Hamiltonian Methods in Field Theory} (World Scientific,
Singapore, 1997).

\bibitem{lmp} G.Giachetta, L. Mangiarotti and G. Sardanashvily,
Iterated BRST cohomology {\it Lett. Math. Phys.} {\bf 53} (2000) 143.

\bibitem{jmp} G.Giachetta, L. Mangiarotti and G. Sardanashvily, 
Cohomology of the infinite-order jet space and the inverse problem, 
{\it J. Math. Phys.} {\bf 42} (2001) 4272.

\bibitem{got} M.Gotay, A multisymplectic framework for classical field
theory and the calculus of variations, I. Covariant Hamiltonian formalism,
In: {\it Mechanics, Analysis and Geometry; 200 Years after Lagrange}
(North Holland, Amsterdam, 1991) 203.

\bibitem{hir}  F.Hirzebruch, {\it Topological Methods in Algebraic
Geometry} (Springer-Verlag, Berlin, 1966). 

\bibitem{krup1} D.Krupka, Variational sequences and variational
bicomplexes, In: {\it Proc. VII Conf. Dif. Geom. Appl., Satelite Conf.
of ICM in Berlin (Brno, 1998)} (Masaryk Univ., Brno, 1999) 525.

\bibitem{krup2} D.Krupka and J.Musilova, Trivial Lagrangians in field
theory, {\it Diff. Geom. Appl.} {\bf 9} (3) (1998) 293.

\bibitem{krup3} D.Krupka, On the local structure of the Euler--Lagrange
mapping of the calculus of variations, {\it E-print arXiv}:
math-ph/0203024.

\bibitem{olver} P.Olver, {\it Applications of Lie Groups to
Differential Equations} (Springer-Verlag, Berlin, 1997). 

\bibitem{mpla} G.Sardanashvily, Cohomology of the variational complex
in field-antifield BRST theory, {\it Mod. Phys. Lett. A} {\bf 16}
(2001) 1531; {\it E-print arXiv}: hep-th/0102175.

\bibitem{ijmms} G.Sardanashvily, Cohomology of the variational complex
in the class of exterior forms of finite jet order, {\it Int. J. Math.
and Math. Sci.} {\bf 30} (2002) 39.

\bibitem{epr} G.Sardanashvily, Ten lectures on jet manifold in
classical and quantum field theory, {\it E-print arXiv}: math-ph/0203040.

\bibitem{tak1}  F.Takens, Symmetries, conservation laws and variational
principles, In: {\it Geometry and Topology}, Lect. Notes in Mathematics
{\bf 597} (Springer-Verlag, Berlin, 1977) 581. 

\bibitem{tak2} F.Takens, A global version of the inverse problem
of the calculus of variations, {\it J. Diff. Geom.} {\bf 14} (1979) 543. 

\bibitem{tul}  W.Tulczyjew, The Euler--Lagrange resolution, 
In: {\it Differential
Geometric Methods in Mathematical Physics}, Lect. Notes in Mathematics 
{\bf 836} (Springer-Verlag, Berlin, 1980) 22. 

\bibitem{vin} A.Vinogradov, The $\cal C$-spectral sequence, Lagrangian
formalism and conservation laws. II The nonlinear theory, {\it J. Math.
Anal. Appl.} {\bf 100} (1984) 41.

\bibitem{vit} R.Vitolo, Finite order Lagrangian bicomplex, {\it Math.
Proc. Cambridge Phil. Soc.} {\bf 125} (1998) 321.

\end{thebibliography}
\end{document}